\documentclass[sigconf]{acmart}

\copyrightyear{2025} 
\acmYear{2025} 
\setcopyright{cc}
\setcctype{BY}
\acmConference[KDD '25]{Proceedings of the 31st ACM SIGKDD Conference on Knowledge Discovery and Data Mining V.1}{August 3--7, 2025}{Toronto, ON, Canada}
\acmBooktitle{Proceedings of the 31st ACM SIGKDD Conference on Knowledge Discovery and Data Mining V.1 (KDD '25), August 3--7, 2025, Toronto, ON, Canada}
\acmDOI{10.1145/3690624.3709284}
\acmISBN{979-8-4007-1245-6/25/08}

\makeatletter
\gdef\@copyrightpermission{
  \begin{minipage}{0.3\columnwidth}
   \href{https://creativecommons.org/licenses/by/4.0/}{\includegraphics[width=0.90\textwidth]{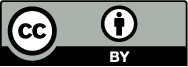}}
  \end{minipage}\hfill
  \begin{minipage}{0.7\columnwidth}
   \href{https://creativecommons.org/licenses/by/4.0/}{This work is licensed under a Creative Commons Attribution International 4.0 License.}
  \end{minipage}
  \vspace{5pt}
}
\makeatother

\usepackage{booktabs}
\usepackage{boldline} 
\usepackage{multirow}
\usepackage{colortbl}
\usepackage{xcolor}
\usepackage{array}
\usepackage{cleveref}
\usepackage{algorithm}
\usepackage{algorithmic}
\usepackage{balance}
\usepackage{pifont}
\usepackage{graphicx}
\usepackage{subcaption} 

\begin{document}

\title{Breaking the Memory Wall for Heterogeneous Federated Learning via Progressive Training}


\author{Yebo Wu}
\orcid{0000-0002-2422-1356}
\affiliation{%
    \institution{University of Macau}
    \department{State Key Laboratory of IoTSC}
    \city{Macau SAR}
    \country{China}}
\email{yc37926@um.edu.mo}

\author{Li Li}
\authornote{The corresponding author.}
\orcid{0000-0002-2044-8289}
\affiliation{%
    \institution{University of Macau}
    \department{State Key Laboratory of IoTSC}
    \city{Macau SAR}
    \country{China}}
\email{llili@um.edu.mo}

\author{Cheng-zhong Xu}
\orcid{0000-0001-9480-0356}
\affiliation{%
    \institution{University of Macau}
    \department{State Key Laboratory of IoTSC}
    \city{Macau SAR}
    \country{China}}
\email{czxu@um.edu.mo}

\renewcommand{\shortauthors}{Yebo Wu, Li Li, \& Cheng-zhong Xu}


\begin{abstract}
Federated Learning (FL) enables multiple devices to collaboratively train a shared model while preserving data privacy. Most existing research assumes that all participating devices have sufficient resources to support the training process. However, the high memory requirements of model training present a significant challenge to deploying FL on resource-constrained devices in practical scenarios. To this end, this paper presents \textbf{ProFL}, a new framework that effectively addresses the memory constraints in FL. Rather than updating the full model during local training, ProFL partitions the model into blocks based on its original architecture and trains each block in a progressive fashion. It first trains the front blocks and safely freezes them after convergence. Training of the next block is then triggered. This process progressively grows the model to be trained until the training of the full model is completed. In this way, the peak memory footprint is effectively reduced for feasible deployment on heterogeneous devices. In order to preserve the feature representation of each block, the training process is divided into two stages: model shrinking and model growing. During the model shrinking stage, we meticulously design corresponding output modules to assist each block in learning the expected feature representation and obtain the initialization model parameters. Subsequently, the obtained output modules and initialization model parameters are utilized in the corresponding model growing stage, which progressively trains the full model. Additionally, a novel metric from the scalar perspective is proposed to assess the learning status of each block, enabling us to securely freeze it after convergence and initiate the training of the next one. Finally, we theoretically prove the convergence of ProFL and conduct extensive experiments on representative models and datasets to evaluate its effectiveness. The results demonstrate that ProFL effectively reduces the peak memory footprint by up to 57.4\% and improves model accuracy by up to 82.4\%.
\end{abstract}



\maketitle

\section{Introduction}
\label{sec:intro}

Federated Learning (FL)~\cite{mcmahan2017communication, tian2022harmony, fu2022federated, ye2024praffl} is a new learning paradigm that enables multiple devices to collaboratively train a global model while preserving data privacy. Most existing approaches assume that the participating devices possess sufficient memory capacity to update the global model~\cite{wang2023fedins2, ning2024fedgcs, fu2024federated2, tam2023federated, fu2024federated}. However, in real-world scenarios, the participating devices, such as smartphones and wearable devices, typically have limited resources~\cite{kou2024fast, tian2023learn}. For instance, training ResNet50 on ImageNet requires more than 12 GB of memory. However, commonly used mobile devices offer only 4 to 12 GB of available memory~\cite{tam2024fedhybrid}.
Moreover, to enhance analytical capabilities, models are growing more complex by adding additional layers and parameters~\cite{kou2024pfedlvm}, which in turn leads to increased memory usage.
Such resource constraints prevent many memory-constrained devices from contributing to the global model with their private data~\cite{tian2024breaking, zhan2024heterogeneity}.
On the other hand, using a small model that all devices can support for local training as the global model results in limited representational capacity and significantly narrows the range of FL applications~\cite{liu2022no, wu2024neulite}.

In order to surmount the resource limitation, several works have been proposed and can be divided into the following two categories~\cite{wu2024heterogeneity, wuhoney}: 1) model-heterogeneous FL and 2) partial training. 
The approaches in the first category customize local models for each participating device to match its memory capacity~\cite{li2019fedmd, itahara2021distillation, cho2022heterogeneous, zhang2022fedzkt}. Knowledge distillation~\cite{hinton2015distilling} is then performed for model aggregation. However, a public dataset is required to complete the information transfer between different model architectures, which is usually hard to retrieve due to privacy concerns in real-world scenarios. For the second category, width scaling~\cite{diao2020heterofl, alam2022fedrolex, horvath2021fjord} and depth scaling~\cite{kim2022depthfl, liu2022no} are widely adopted. 
For width scaling, the number of channels in the convolutional layers is scaled to resize the model, thereby accommodating the memory capacity of the devices.
However, this approach substantially compromises the model architecture and simultaneously deteriorates the overall performance. For depth scaling, devices can train models of different depths based on their memory constraints. However, this method has a strong requirement that a certain portion of the participating devices must be capable of training the full model. Thus, the complexity of the global model that can be trained is bounded by these high-end devices. To this end, a new training paradigm that can effectively break the memory wall without requiring a public dataset is crucial for FL in real-world deployment.

In this paper, we propose ProFL, a novel progressive training approach that addresses the memory constraints in FL from a new viewpoint. 
In contrast to standard FL, which updates the full model continuously, ProFL employs a well-designed progressive training paradigm, updating only a portion of the model in each round and progressively completing the training of the full model. ProFL first divides the model into different blocks, allowing devices that can afford the training of the current block to participate. Then, when the current block converges, this block is frozen, and a new block is added for training on top of it. 
In this way, the frozen parts no longer require backpropagation, thereby effectively freeing up memory space previously allocated for storing the activations of these layers. This process iterates until all blocks have been successfully trained. As the peak memory footprint during the training process is notably reduced, devices with constrained memory resources can effectively participate in the training process.

Specifically, ProFL features two stages: 1) progressive model shrinking and 2) progressive model growing. 
Progressive model shrinking is designed to construct corresponding output modules and acquire the initialization parameters for each block. During the training process of each block, except for the last block, which has an output module for end-to-end training, the other blocks lack the ability to undergo independent training. Simply adding a fully connected layer can severely compromise the feature representation of each block, as each block in the model is responsible for extracting different levels of features~\cite{chen2022layer}. To tackle this challenge, we employ basic layers to substitute each subsequent block, mimicking the position of the training block in the original model and preserving their position information. In particular, we first train each block of the model from back to front and integrate the trained block information into a basic layer. Through this process, basic layers carrying block-specific information for building output modules can be retrieved. At the same time, we acquire the initialization parameters for each block, resulting in better optimization results. Then, the retrieved output modules and initialization parameters are utilized in the progressive model growing stage, which sequentially trains each block from front to back till the training of the full model is completed.

Furthermore, accurately evaluating the training progress of each block is crucial. Inappropriately freezing non-converged blocks will compromise model performance because it can prevent the model from fully learning and adapting to the training data, leading to suboptimal feature representation.
To address this challenge, the block freezing determination module is designed to assess each block's training progress, securely freeze it upon convergence, and subsequently initiate the training of the next block. Specifically, we propose a new metric from the scalar perspective, which is used to capture the updated status of each scalar by analyzing its movement distance.
Initially, large gradients drive effective movement toward the optimal solution. As training proceeds, diminishing gradients lead to reduced scalar movement distance. Ultimately, scalar oscillation at the optimal solution results in near-zero movement distance within a window, signifying convergence. We sum movement distances of all scalars within a block, termed effective movement, as a measure of block training progress. This metric facilitates accurate tracking of individual block learning progress, allowing safe initiation of block freezing without compromising model performance.

In summary, our main contributions are as follows: 1) We propose ProFL, a novel FL approach that effectively addresses the memory constraints via progressive block training. We then decouple the model training process into two stages to assist each block in learning the expected feature representation. 2) To precisely evaluate the learning status of each block, we propose a novel metric termed effective movement, viewed from the scalar perspective. 3) We theoretically prove the convergence of ProFL and conduct extensive experiments to validate its superiority on representative models and datasets.
\section{ProFL}

\subsection{High-Level Ideas}

\begin{figure}[!t]
  \centering
  \includegraphics[width=1\linewidth]{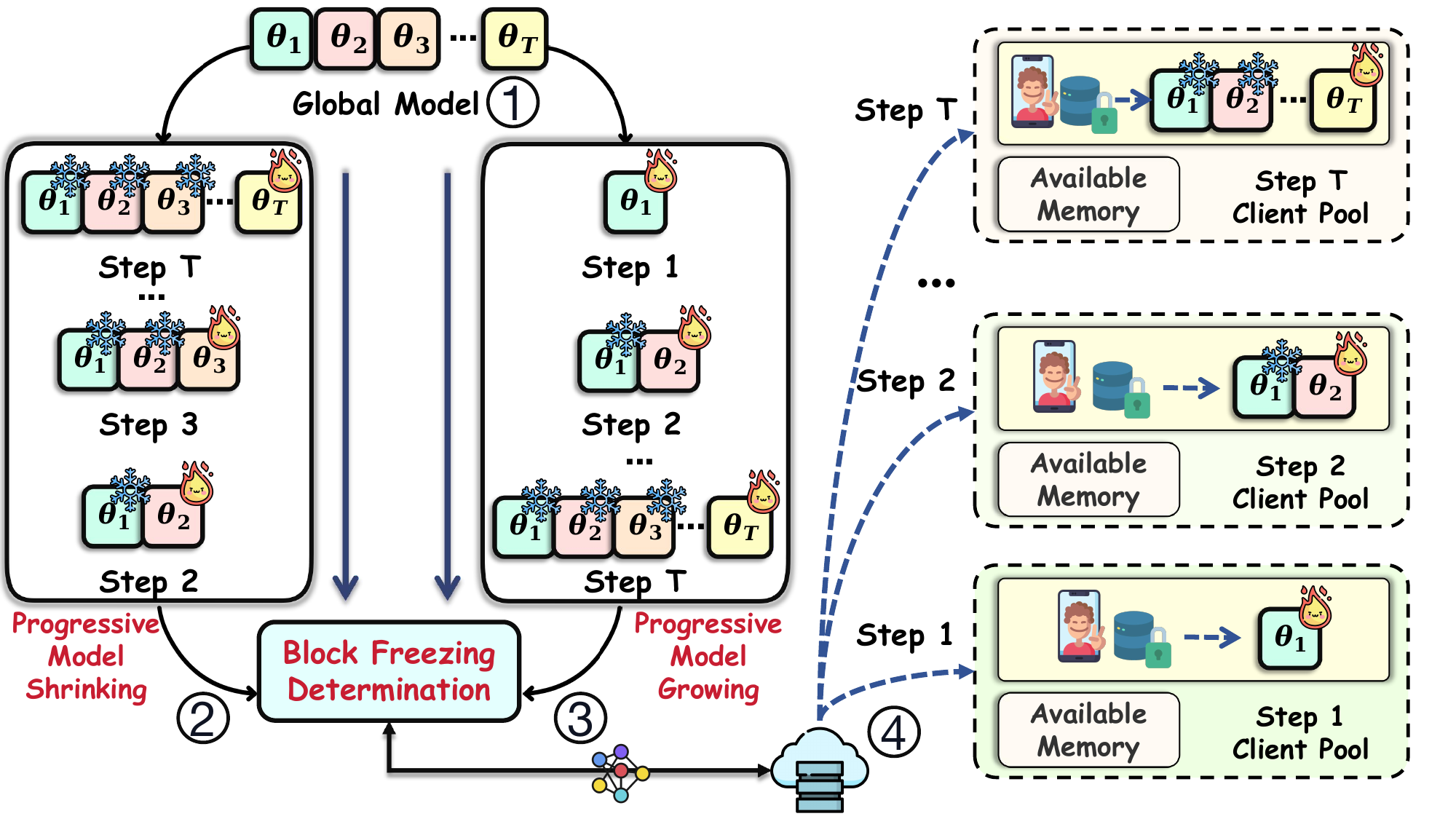}
  \caption{The workflow of ProFL. The global model is divided into multiple blocks. Progressive model shrinking is initially performed, followed by progressive model growing. Block freezing determination is employed to evaluate the training status of each block during both stages.}
  \label{workflow}
\end{figure}

ProFL is designed to efficiently train a global model $\Theta$ under resource-constrained conditions.
Fig.~\ref{workflow} presents the overall workflow of ProFL. ProFL first divides the global model into $T$ ([$\theta_{1},\theta_2,...,\theta_T$]) blocks based on the model architecture (\ding{172} in Fig.~\ref{workflow}), corresponding to different training steps. Subsequently, the training process is triggered. Instead of updating the full model, ProFL trains each block in a progressive fashion (\ding{174} in Fig.~\ref{workflow}). For instance, ProFL initially trains the first block $\theta_{1}$ and safely freezes it after convergence. Training of the second block $\theta_{2}$ is then triggered. This stage gradually expands the model from the first block until each block has undergone sufficient training, ultimately achieving the target model (see Section~\ref{sec:FL_training} for details). 
In this progressive training process, since only the corresponding block is trained, this effectively reduces the peak memory footprint. However, except for the last block $\theta_{T}$, the other blocks cannot independently undergo end-to-end training due to the absence of output modules. To facilitate this progressive training process, we introduce the progressive model shrinking stage (\ding{173} in Fig.~\ref{workflow}), which sequentially trains each block of the model from back to front and integrates the information from each trained block into a basic layer to construct output modules for each block, guiding its learning process (see Section~\ref{sec:model_shrinking} for details). The purpose of this stage is twofold: 1) obtaining the initialization parameters for each block and 2) constructing the corresponding output modules for the progressive model growing stage by utilizing the obtained basic layers. At each step, the client set $S$ is selected from the pool of clients who can afford training for the current block (\ding{175} in Fig.~\ref{workflow}). Throughout both stages, the training pace of each step is controlled by the block freezing determination module to efficiently determine whether the current block has been well trained, securely freeze it upon convergence, and initiate the next training step (see Section~\ref{sec:model_convergence} for details).

\subsection{Progressive Training Paradigm}\label{sec:FL_training}

\begin{figure}[!t]
  \centering
  \includegraphics[width=0.85\linewidth]{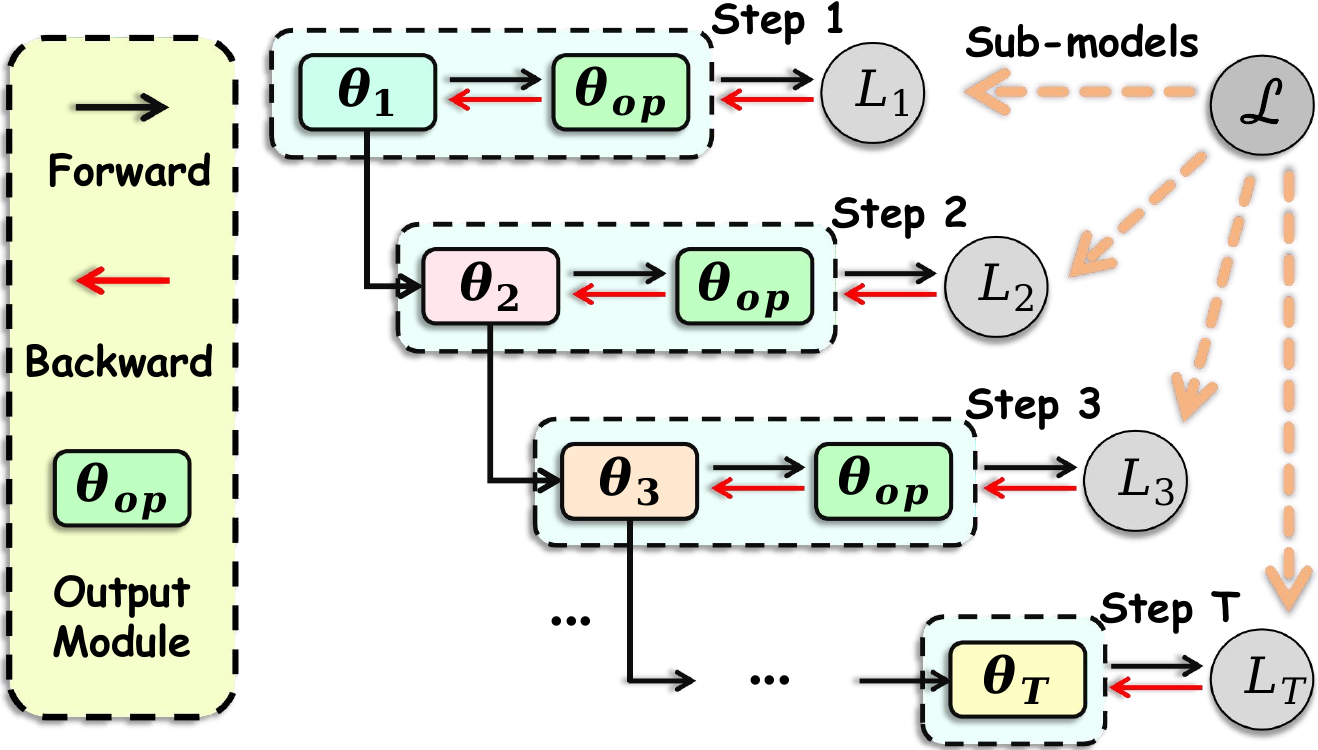}
  \caption{Progressive Model Growing. In each step t, only the corresponding block $\theta_{t}$ and output module $\theta_{op}$ are updated.}
  \label{growing}
\end{figure}

In this section, we introduce the progressive training paradigm, namely, the progressive model growing stage, as shown in Fig.~\ref{growing}.
Assume that $N$ devices participate in an FL system, with each device maintaining a local dataset denoted as $D_n$ (where $n \in [1, N]$).
The global model $\Theta$ to be trained is partitioned into $T$ blocks. The block being actively trained is represented as $\theta_{t}$ ($1 \leq t \leq T$), the frozen block as $\theta_{t,F}$, and the well-trained block as $\theta_{t}^{*}$. The sub-model of the $t$-th step during progressive model growing is denoted as $\Theta_{t} = [\theta_{1,F}^{*},\theta_{2,F}^{*},...,\theta_{t}]$. When $1 \leq t < T$, the sub-models cannot independently undergo end-to-end training due to the absence of output modules. To facilitate this training process, adding an output module $\theta_{op}$ to each training block is essential. Consequently, the model trained at step $t$ can be represented as $\Theta_{t} = [\theta_{1,F}^{*},\theta_{2,F}^{*},...,\theta_{t},\theta_{op}]$. Once the current training block converges, it advances to step $t+1$. 
This process repeats until the full model has been well optimized.

In this step-wise progressive training paradigm, each round is performed as follows: 1) The central server constructs the global sub-model $\Theta_{t}^{r} ([\theta_{1,F}^{*},\theta_{2,F}^{*},..., \theta_{t},\theta_{op}])$ of the round $r$ to be trained; 2) The central server then selects the client set $S$ based on their available memory and sends $\Theta_{t}^{r}$ to them; 3) The selected clients train $\Theta_{t}^{r}$ on their local datasets. Since the previous blocks are frozen, only parameters $[\theta_{t}, \theta_{op}]$ are updated; 4) After completing local training, the updated parameters $[\theta_{t,n}^{r}, \theta_{op,n}^{r}]$ (where $n \in S$) are uploaded to the central server; 5) The central server subsequently aggregates the updates according to Eq.~\eqref{eq_aggre} to obtain the global model for the next round, $\Theta_{t}^{r+1}$.
\begin{equation}
    \label{eq_aggre}
    [\theta_{t,g}^{r+1}, \theta_{op,g}^{r+1}] = \sum_{n \in S} \frac{|D_{n}|}{|D|}([\theta_{t,n}^{r}, \theta_{op,n}^{r}])
\end{equation}
where $|D_{n}|$ represents the size of local dataset on client $n$, $|D|$ represents the amount of the overall training data, $[\theta_{t,n}^{r},\theta_{op,n}^{r}]$ stands for the updated model parameters of client $n$ on round $r$, and $[\theta_{t,g}^{r+1}, \theta_{op,g}^{r+1}]$ represents the aggregated global model parameters. Following parameter aggregation, the block freezing determination module assesses the training progress of the current block. If the current block is well-trained, it safely freezes this block and triggers the next training step. Upon convergence of the current block, we obtain $\theta_{t}^{*}$. The model for the next step $t+1$ is derived by introducing a new block $\theta_{t+1}$ based on the blocks trained in the preceding $t$ steps, represented as $\Theta_{t+1} = [\theta_{1,F}^{*},\theta_{2,F}^{*},...,\theta_{t,F}^{*},\theta_{t+1},\theta_{op}]$.

However, the progressive training paradigm encounters two primary challenges. The first challenge revolves around constructing corresponding output modules $\theta_{op}$ to aid each block in learning the expected feature representation. Inappropriately constructed output modules can significantly impact the final model performance as they can bias the training process, affecting the feature representation learned by the blocks. To address this challenge, progressive model shrinking is designed to retrieve the basic layers with block-specific information for constructing output modules and acquiring the initialization parameters for each block. After that, the corresponding progressive model growing stage can be conducted. The second challenge pertains to controlling the training pace of each block. As highlighted in \cite{wang2023egeria}, improper freezing may adversely affect the final model performance as non-converged blocks cannot sufficiently learn and adapt to the training data. To this end, we introduce a novel metric from the scalar perspective to assess the training progress of each block, ensuring judicious freezing for optimal results.

\begin{figure}[!t]
  \centering
  \includegraphics[width=1\linewidth]{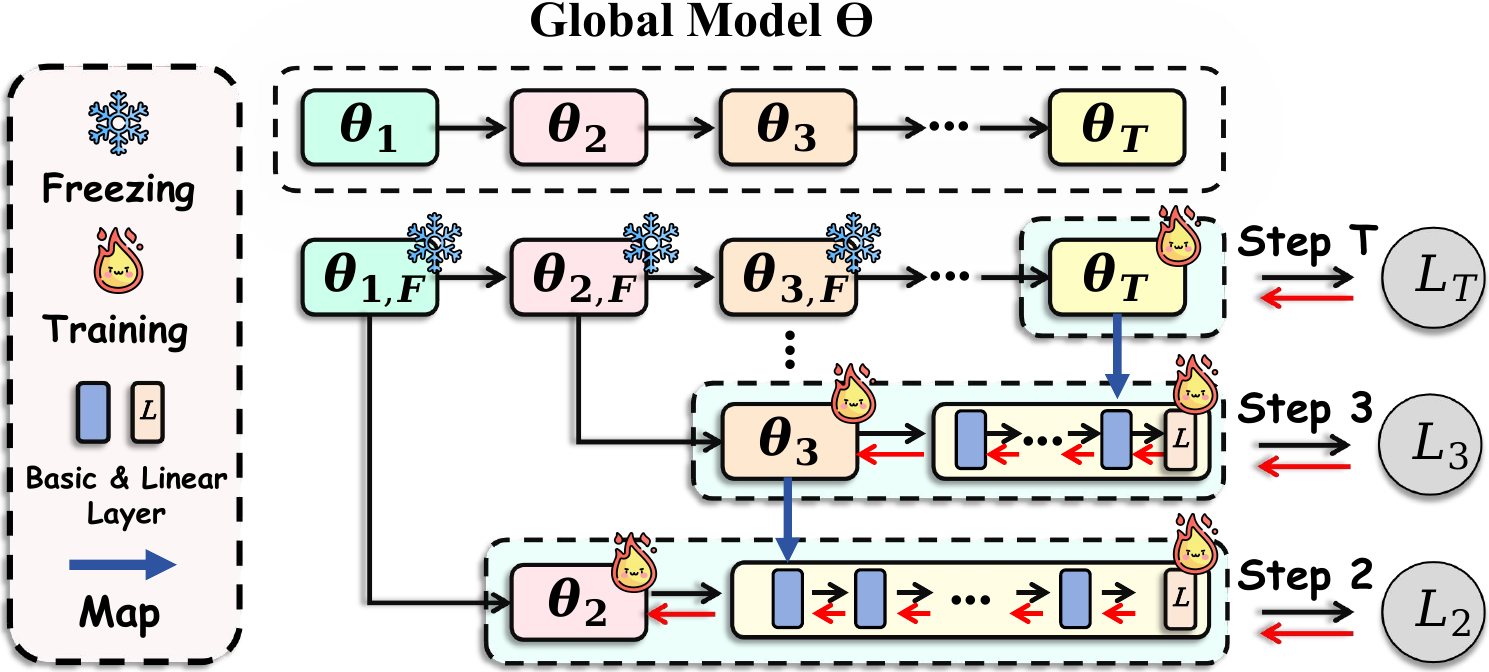}
  \caption{Progressive Model Shrinking. \textit{Map} implies integrating the information learned by the block into a basic layer.}
  \label{graph_framework}
\end{figure}

\subsection{Progressive Model Shrinking}\label{sec:model_shrinking}

Fig.~\ref{graph_framework} shows the overall workflow of progressive model shrinking. It transforms the model in the opposite direction, serving two primary purposes. Firstly, the parameters obtained by training each block from back to front are utilized to initialize the corresponding block in the progressive model growing stage. Secondly, following the training of each block, information from the block is integrated into a basic layer, such as a convolutional layer, which acts as a component of the output module, guiding blocks in learning their expected feature representation. For example, as shown in Fig.~\ref{graph_framework}, after completing the training of  $\theta_{T}$, the information from this block is integrated into a basic layer. Information integration is completed through knowledge distillation~\cite{hinton2015distilling}. In conjunction with a fully connected layer, the resulting basic layers collectively form the output module $\theta_{op}$. Specifically, the basic layer corresponding to block $t$ is denoted as $\theta_{t,b}$, and the fully connected layer is denoted as $\theta_{L}$. As illustrated in Fig.~\ref{graph_framework}, the output module used in each step of progressive model shrinking corresponds to the one applied in each step of progressive model growing.

The progressive model shrinking stage primarily includes the following key steps: 1) For step $t$ during progressive model shrinking, the sub-model is represented as $\Theta_{t}=[\theta_{1,F},\theta_{2,F},...,\theta_{t},\theta_{t+1,b},...,\theta_{T,b},\theta_{L}]$, undergoing continuous training until block $\theta_{t}$ converges, yielding $\theta_{t}^{ini}$. 
It is worth noting that, during the training process of each step, only the current block and the output module are updated. This significantly reduces the computational load of both forward and backward propagation, minimizing training overhead and effectively reducing peak memory footprint. 2) Fig.~\ref{graph_framework} illustrates that, upon convergence, the information from block $\theta_{t}^{ini}$ is integrated into a basic layer $\theta_{t,b}$, and a new output module $\theta_{op}$ ($[\theta_{t,b},\theta_{t+1,b},...,\theta_{T,b},\theta_{L}]$) for model $\Theta_{t-1}$ is constructed. Subsequently, it proceeds to step $t-1$ and initiates the training of $\theta_{t-1}$. These two steps are iteratively performed until $\theta_{2}$ completes training and integrates its information into the corresponding basic layer. After this stage, we can obtain the initialization parameters for each block and basic layers with block-specific information to construct the output modules.

\subsection{Block Freezing Determination}\label{sec:model_convergence}

The block freezing determination module is hosted on the server, where it evaluates the learning progress of each block, securely freezes it upon convergence, and initiates the training process for the next block.
Layer freezing~\cite{wang2023egeria, li2022smartfrz} is widely adopted in centralized training to expedite the model training process and reduce computational overhead. However, these methods are not directly applicable to the FL scenario. For example, SmartFRZ~\cite{li2022smartfrz} utilizes training histories to develop an attention-based predictor, which is then used to estimate the freezing probability of a layer during the learning process. However, retrieving such histories is challenging due to data privacy issues.
To effectively monitor the learning status of each block, we propose an innovative metric from the scalar perspective. This metric guides both progressive model shrinking and progressive model growing stages.

Specifically, we define the update of a scalar $s$ at the $k$-th epoch as $\epsilon_{s}^{k}=s^{k}-s^{k-1}$, and $||\epsilon_{s}^{k}||$ represents the movement distance of the scalar. Consequently, the absolute movement distance of this scalar over $H$ consecutive epochs can be expressed as $D_{s,k}^{H}=||\sum_{h=0}^{H-1} \epsilon_{s}^{k-h}||$. As a parameter tends to converge, $D_{s,k}^{H}$ approaches 0. The underlying reason is that when the gradient is relatively large at the outset, the scalar swiftly moves in the direction of the optimal solution, and this movement direction remains consistent within the window size $H$, thus resulting in a relatively large absolute movement distance. As training proceeds, the gradient decreases, leading to a reduction in the movement distance. When the scalar is close to the optimal solution, it oscillates near this point, causing the movement distance within the window $H$ to approach 0.

The movement distance of all parameters within a block can be expressed as $D_{B,k}^{H}=\sum_{s \in B} D_{s,k}^{H}$, indicating the overall update of the block. A larger $D_{B,k}^{H}$ signifies that all scalars in the block are actively moving towards the optimal point. Conversely, if $D_{B,k}^{H}$ is smaller, it suggests that all parameters in the block are near the optimum point, indicating that the block is approaching convergence. Considering the varying parameter quantities and update scales across different blocks and different scalars, we normalize $D_{B,k}^{H}$ by dividing it by $\sum_{s \in B}\sum_{h=0}^{H-1} ||\epsilon_{s}^{k-h}||$. The resulting normalized value $\frac{D_{B,k}^{H}}{\sum_{s \in B}\sum_{h=0}^{H-1} ||\epsilon_{s}^{k-h}||}$ is defined as effective movement, providing insight into the learning status of the block. Therefore, we evaluate the training progress of each block by calculating its effective movement. 
Subsequently, we apply linear least-squares regression~\cite{watson1967linear} to model the effective movement and analyze the slope to determine the optimal freezing time. If the slope is below threshold $\phi$ for $W$ times, we freeze it and proceed to the next step.

\subsection{Convergence Analysis}\label{sec:model_analysis}

In this section, we prove the convergence of the freezing-based progressive training paradigm based on~\cite{haddadpour2019convergence, li2019convergence}. While prior works have demonstrated the convergence of progressive training~\cite{mohtashami2021simultaneous}, the key difference lies in their neglect of memory constraints. Consequently, in proving convergence, they solely focus on assessing the similarity of gradients between sub-models and the global model. Unlike them, we freeze the well-trained blocks at each step. For example, in step $t$, we exclusively train the block $\theta_{t}$ of the model $\Theta_{t}$. Without loss of generality, we omit the output module here, while considering the case of full device participation in each step.

\noindent
\textbf{Notation.} Let $f_1, \ldots, f_N$ represent local functions, $M_{t}$ indicate total SGDs performed by each device at step $t$, and $E$ be the number of local epochs. Assume $\mathbf{x}$ and $\mathbf{y}$ are any two points in the function's domain. When training neural networks, our goal is to find the optimal global model parameters $\Theta^{*}$ that minimize the empirical loss $f$ ($\mathbb{R}^{d} \rightarrow \mathbb{R}$): $f^{*} := \min_{\Theta \in \mathbb{R}^{d}} f(\Theta)$.

\noindent
\textbf{Assumption 1} ($\mu$-strongly convex). For any arbitrary $\textbf{x}$ and $\textbf{y}$, there exists:
\begin{equation}
    f_n(\textbf{x}) - f_n(\textbf{y}) \geq (\textbf{x}-\textbf{y})^{T}\nabla f_{n}(\textbf{y}) + \frac{\mu}{2}||\textbf{x}-\textbf{y}||_{2}^{2}, \forall~ \textbf{x},\textbf{y}, n
    \label{convex}
\end{equation}
\noindent
\textbf{Assumption 2} (L-smooth). The functions $f_1, ..., f_N$: $\mathbb{R}^{d} \to \mathbb{R}$ is differentiable and there exists a constant $L > 0$ such that:
\begin{equation}
    f_n(\textbf{x}) - f_n(\textbf{y}) \leq (\textbf{x}-\textbf{y})^{T}\nabla f_{n}(\textbf{y}) + \frac{L}{2}||\textbf{x}-\textbf{y}||_{2}^{2}, \forall~ \textbf{x},\textbf{y}, n
    \label{lsmooth}
\end{equation}
This ensures that the gradient cannot change arbitrarily fast, a weak assumption that is practically guaranteed in most machine learning models~\cite{li2022one, haddadpour2019convergence}

\noindent
\textbf{Assumption 3} (Bounded stochastic gradient variance). Let $\xi_{t,n}^{m}$ be sampled from the $n-$th device's local data uniformly at random. For any $\Theta_{t,n}^{m} \in \mathbb{R}^{d}$, we can obtain bounded stochastic gradient variance~\cite{zhang2012communication, yu2019parallel, stich2018local}. For any $t \in \{1,2,...,T\}$ and $m \in \{1,2,...,M_{t}\}$, there exists parameters $\sigma_{t,n}^{2} \geq 0 $ that satisfies:
\begin{equation}
    \mathbb{E}||\nabla f_{n}(\Theta_{t,n}^{m},\xi_{t,n}^{m}) - \nabla f_n(\Theta_{t,n}^{m})||^{2} \leq \sigma_{t,n}^{2}, \forall~ t, n, m
    \label{bound}
\end{equation}
When $t=1$, it signifies that an individual model is undergoing end-to-end training.
This assumption translates into the standard assumption used for analyzing the convergence of FedAvg under data heterogeneity~\cite{li2019convergence}. When $t \neq 1$, the step-wise sub-model $\Theta_{t}$ is constructed by adding $\theta_{t}$ to the model $\Theta_{t-1}$.

\noindent
\textbf{Assumption 4} (Uniform stochastic gradient). The expected squared norm of stochastic gradients is uniformly bounded~\cite{haddadpour2019convergence, stich2018local, stich2018sparsified}, i.e., $\mathbb{E}||\nabla f_n(\Theta_{t,n}^{m},\xi_{t,n}^{m})||^{2} 
\leq G^2$ for all $n =1,...,N$, $t=1,...,T$ and $m = 1,...,M_{t}-1$.

Assumption 1 formally establishes a fundamental lower bound crucial for ensuring the convergence of the loss function to a stationary point. It implies that a globally optimal solution $\Theta^{*}$ is guaranteed to exist. Assumptions 2-4 are standard stochastic optimization assumptions~\cite{haddadpour2019convergence, stich2018local, li2019convergence}. Utilizing the above standard assumptions, we first prove the convergence of the model at $t=1$ and then extend it to the model at $t \neq 1$.

\noindent
\textbf{Theorem 1.}\label{theorem} Let Assumptions 1, 2, 3, and 4 hold, and let $L, \mu, \sigma_{t,n}, G$ be as defined therein. Let $\kappa = \frac{L}{\mu}$, $\gamma = \max\{8\kappa, E\}$ and the stepsize $\eta_{m} = \frac{2}{\mu(\gamma+m)}$. Assuming $f^{*}(\Theta_{t})$ and $f_{n}^{*}(\Theta_{t})$ are the minimum values of $f(\Theta_{t})$ and $f_{n}(\Theta_{t})$ respectively, and $p_{n}$ is the aggregation weight for device $n$, the data heterogeneity degree is expressed as $\Gamma = f^{*}(\Theta_{t}) - \sum_{n=1}^{N}p_{n}f_{n}^{*}(\Theta_{t})$. Then:
\begin{align}
    \mathbb{E}[f(\textbf{$\Theta_{t}^{M_{t}}$})] - f^{*}(\Theta_{t}) &\leq \frac{\kappa}{\gamma+M_{t}-1}\left(\frac{2B}{\mu}+\frac{\mu \gamma}{2}\mathbb{E}\|\textbf{$\Theta_{t}^{1}$}-\textbf{$\Theta_{t}^{*}$}\|^2\right)
    \label{convergence1}
\end{align}
where $B = \sum_{n=1}^{N}p_{n}^{2}\sigma_{t,n}^{2}+6L\Gamma+8(E-1)^{2}G^{2}$.
It can be seen that, when $t$ is ignored, for strongly convex and smooth functions under data heterogeneity, 
FedAvg achieves convergence to the global optimum with a rate of $\mathcal{O}(\frac{1}{M_{t}})$.
When $t=1$, Eq.~\eqref{convergence1} simplifies to the standard convergence proof for end-to-end model training in FL~\cite{li2019convergence}. When $t\neq 1$, due to the freezing of the preceding blocks, their exclusive function is to map the data from a low-dimensional space to a high-dimensional space. Therefore, training the subsequent blocks is still equivalent to training an end-to-end model. Hence, ProFL's convergence is not impacted by block division and the pace-control metric. The difference between this freezing-based progressive training framework and end-to-end training is that model training is broken down into multiple end-to-end trainings. Employing this progressive training paradigm based on freezing, the maximum number of iterations required for the full model to converge is, at most, $T$ times the iterations needed for the direct training of the full model. 
It is worth noting that although the total number of iterations increases, the per-iteration computational overhead during sub-model training is significantly reduced.

\section{Experiments}

\subsection{Experimental Settings}
\textbf{Datasets and Models.} We evaluate the effectiveness of ProFL on the CIFAR10 and CIFAR100 benchmark datasets~\cite{krizhevsky2009learning} using representative model architectures including ResNet18, ResNet34~\cite{he2016deep}, VGG11$\_$bn, and VGG16$\_$bn~\cite{simonyan2014very}.
Additionally, to demonstrate the versatility of ProFL, we further conduct experiments with Vision Transformer (ViT)~\cite{dosovitskiy2020image} on Tiny-ImageNet~\cite{le2015tiny} and evaluate its scalability on the large-scale FEMNIST dataset~\cite{caldas2018leaf}.
For the VGG11$\_$bn model, we add a max-pooling layer for downsampling after every two convolutional layers. For the VGG16$\_$bn model, we add a max-pooling layer for downsampling after every four convolutional layers. 
The datasets are distributed in both IID and Non-IID settings~\cite{tian2024ranking}, where the Non-IID follows a Dirichlet distribution with a concentration parameter $\alpha = 1$.

\noindent
\textbf{Baselines.} We compare the performance of ProFL with the following baselines:
\begin{itemize}
    \item \textit{AllSmall}~\cite{liu2022no}: A straightforward method that adjusts the channel count in convolutional layers of the global model
    according to the device with the minimum memory capacity, yielding a model that enables all devices to participate in the FL process.
    \item \textit{ExclusiveFL}~\cite{liu2022no}: This method only allows devices with sufficient memory to participate in FL, which hinders low-memory devices from contributing to the global model.
    \item \textit{HeteroFL}~\cite{diao2020heterofl}: A width scaling method that scales the convolutional layers of the global model based on the memory capacity of each device, creating local models of varying complexities to accommodate memory constraints.
    \item \textit{DepthFL}~\cite{kim2022depthfl}: A depth scaling method that addresses memory constraints by generating sub-models with varying depths. Although both \textit{DepthFL} and \textit{InclusiveFL}~\cite{liu2022no} leverage depth-based model scaling, \textit{DepthFL} achieves superior performance and thus serves as our primary baseline for comparison.
\end{itemize}

\noindent
\textbf{Default Settings.} We establish a pool of 100 devices and randomly select 20 devices in each training round for the image classification tasks. 
To maintain fairness in comparisons, we randomly assign available memory to each device within a range of 100-900 MB~\cite{wu2024heterogeneity}, taking into account resource competition.
During local training, each device performs five local epochs using SGD with a learning rate of 0.01, except for Vision Transformer, which employs AdamW~\cite{loshchilov2017decoupled} with a learning rate of 0.0001 as the optimizer.
For ResNet18 and ResNet34, we divide the models into four blocks corresponding to 4 steps based on the residual blocks. For VGG11$\_$bn, we divide the model into two blocks corresponding to 2 steps, considering the first four convolutional layers and the last four convolutional layers separately. 
For the VGG16$\_$bn model, we divide it into three blocks, with each block corresponding to 4, 4, and 5 convolutional layers, respectively. 
If a device's memory is insufficient to train any block, it is assigned to train the output layer, ensuring full utilization of its valuable data.
In extreme scenarios where the memory capacity of all devices is severely limited, the process transitions to layer-wise training. 
For the selection of other hyperparameters, we set the window size $H$ to 10 or 20, the slope threshold $\phi$ to 10\%-20\% of the initial slope of each block, and $W$ to a value within [20, 40].

\begin{table*}[!h]
  \centering
  \small
  \caption{Performance comparison of FL methods on ResNet and VGG series models. Boldface indicates the best results. $VGG11$ and $VGG16$ refer to the VGG11$\_$bn and VGG16$\_$bn models, respectively. $PR$ denotes the participation rate of the device, and NA signifies that the corresponding algorithm cannot operate under this setup due to memory limitations.}
  \label{performance_acc1}
  \resizebox{\linewidth}{!}{

  \begin{tabular}{c c cccc  cccc cccc}
    \toprule[1pt]

    \multirow{2}{*}{Dataset} & \multirow{2}{*}{Method} & \multicolumn{4}{c}{IID} & \multicolumn{4}{c}{Non-IID} & \multicolumn{4}{c}{PR} \\
    \cmidrule(lr){3-6} \cmidrule(lr){7-10} \cmidrule(lr){11-14}
    && Res18 & Res34 & VGG11 & VGG16 & Res18 & Res34 & VGG11 & VGG16 & Res18 & Res34 & VGG11 & VGG16 \\
    \midrule
    \multirow{5}{*}{CIFAR10} 
    & AllSmall & 76.7\% & 66.9\% &82.1\%&78.8\%& 69.2\% & 53.9\% &75.3\% & 69.8\% & 100\% & 100\% & 100\% & 100\%  \\
    & ExclusiveFL & 65.3\% & NA &83.7\%&NA& 58.6\% & NA & 81.1\% & NA & 8\% & 0\% & 24\% & 0\% \\
    & HeteroFL & 75.5\% & 9.8\% &83.9\%&11.6\%&  62.9\% & 9.6\% & 78.2\% & 10.8\% & 100\% & 100\% & 100\% & 100\% \\
    & DepthFL & 70.4\% & 71.7\% &86.4\%&76.9\%& 60.8\% & 55.9\% & 83.4\% & 71.2\% & 47\% & 34\% & 43\% & 37\%  \\
    & ProFL & \textbf{84.1\%} & \textbf{82.2\%} & \textbf{87.6\%}&\textbf{82.4\%}&\textbf{78.4\%} & \textbf{74.2\%} & \textbf{85.4\%} & \textbf{75.1\%} & {100\%} & {100\%} & 100\% & 100\% \\
    \midrule
    \multirow{5}{*}{CIFAR100} 
    & AllSmall & 37.5\% & 27.3\% & 50.1\% & 38.9\% & 17.1\% & 9.5\% &42.7\% & 31.2\% & 100\% & 100\% & 100\% & 100\% \\
    & ExclusiveFL & 25.7\% & NA & 51.6\% & NA & 23.4\% & NA & 47.5\% & NA & 8\% & 0\% & 24\% & 0\% \\
    & HeteroFL & 36.4\% & 1.3\% & 54.8\% & 1.5\% & 28.1\% & 1.1\% & 47.4\% & 1.3\% & 100\% & 100\% & 100\% & 100\% \\
    & DepthFL & 37.7\% & 28.6\% & 56.1\% & 40.5\% & 26.5\% & 11.5\% & 53.8\% & 36.8\% & 47\% & 34\% & 43\% & 37\% \\
    & ProFL & \textbf{55.4\%} & \textbf{52.3\%} & \textbf{62.9\%} & \textbf{48.4\%} & \textbf{48.3\%} & \textbf{46.1\%} & \textbf{58.8\%} & \textbf{43.8\%} & 100\% & 100\% & 100\% & 100\% \\
    \bottomrule[1pt]
  \end{tabular}}
\end{table*}

\begin{table*}[!ht]
  \centering
  \small
  \caption{Accuracy of step-wise sub-models and global models with and without progressive model shrinking under both IID and Non-IID data distributions on ResNet18 and ResNet34. $PMS$ represents progressive model shrinking.}
  \label{tab_shrinking}
  \resizebox{\linewidth}{!}{
  \begin{tabular}{ccccc cccc cccc c}
    \toprule[1pt]
    \multirow{2}{*}{Distribution} & \multirow{2}{*}{Dataset} & \multirow{2}{*}{PMS} & \multicolumn{2}{c}{Step1} & \multicolumn{2}{c}{Step2} & \multicolumn{2}{c}{Step3} & \multicolumn{2}{c}{Step4} & \multicolumn{2}{c}{Global Model} \\
    \cmidrule(lr){4-5} \cmidrule(lr){6-7} \cmidrule(lr){8-9} \cmidrule(lr){10-11} \cmidrule(lr){12-13}
    & & & Res18 & Res34 & Res18 & Res34 & Res18 & Res34 & Res18 & Res34 & Res18 & Res34 \\
    \midrule
    \multirow{4}{*}{IID} 
    & \multirow{2}{*}{CIFAR10} & \ding{51} & 80.3\% & 79.3\% & 83.4\%&81.7\% & 84.3\% & 82.3\% & 84.6\% & 82.5\% & \textbf{84.1\% (+3.1\%)} & \textbf{82.2\% (+1.4\%)} \\
    & & \ding{55} & 78.3\% & 78.5\% & 80.4\% & 79.9\% &81.1\% & 80.8\% & 81.2\% &81.2\% & 81.0\% & 80.8\%\\
    
    & \multirow{2}{*}{CIFAR100} & \ding{51} & 48.1\% & 48.2\% & 52.9\% & 51.1\% & 55.3\% &51.7\% & 55.9\% &52.5\%& \textbf{55.4\% (+2.8\%)} & \textbf{52.3\% (+3.0\%)} \\
    & & \ding{55} & 41.4\% &44.1\%& 47.5\%&47.6\% & 51.9\% & 49.0\% &52.9\% &49.7\%& 52.6\% &49.3\% \\
    \midrule
    \multirow{4}{*}{Non-IID} 
    & \multirow{2}{*}{CIFAR10} & \ding{51} & 74.9\% &73.1\%& 77.8\%&75.6\% & 79.1\% &75.7\%& 79.1\% &76.1\%& \textbf{78.4\% (+4.7\%)}&\textbf{74.2\% (+2.8\%)} \\
    & & \ding{55} & 71.4\% &71.0\%& 73.4\%&73.1\% & 74.6\% &73.6\%& 74.9\%&74.6\% & 73.7\% & 71.4\%\\
    & \multirow{2}{*}{CIFAR100} & \ding{51} & 44.3\% &43.6\%& 46.8\% &45.1\%& 48.4\%&45.6\% & 48.6\% & 46.6\%&\textbf{48.3\% (+1.4\%)} &\textbf{46.1\% (+0.9\%)} \\
    & & \ding{55} & 38.6\% &39.9\%& 43.5\%&43.8\% & 46.7\% &45.1\%& 47.3\% & 45.6\%&46.9\% &45.2\%\\
    \bottomrule[1pt]
  \end{tabular}}

\end{table*}

\subsection{Performance Evaluation}

Table~\ref{performance_acc1} presents a comprehensive performance comparison of various algorithms on representative datasets and models, evaluated under both IID and Non-IID settings.

For the ResNet18 model, due to its moderate architectural complexity, a subset of devices possesses sufficient memory resources to accommodate the full model training. Specifically, on the CIFAR10 dataset under the IID condition, ProFL surpasses AllSmall by 7.4\% in accuracy. This performance gain is due to AllSmall’s global model being restricted by the device with the minimum available memory, which leads to a simplified model with limited feature extraction capabilities.
Compared to ExclusiveFL, ProFL demonstrates an 18.8\% improvement in accuracy. 
This is because ExclusiveFL only allows devices with sufficient memory to participate, which leads to a mere 8\% participation rate and fails to fully leverage the valuable data from all devices.
In contrast, ProFL significantly reduces peak memory footprint through its progressive training paradigm, facilitating effective model training even on resource-constrained devices.
Additionally, ProFL achieves an 8.6\% performance improvement over HeteroFL. This is because HeteroFL statically partitions convolutional layer channels, assigning different devices to train specific channels, which results in uneven parameter training and compromises the model’s architecture.
As a result, the accuracy of HeteroFL is even lower than that of AllSmall.
Moreover, compared to DepthFL, ProFL still achieves a 13.7\% accuracy improvement. This is because DepthFL overlooks the fact that memory footprint is predominantly caused by earlier blocks, which leads to the exclusion of numerous devices unable to train the first block, resulting in a participation rate of only 47\%. 
Furthermore, as earlier blocks process richer data, only a few devices can train later blocks, leading to uneven parameter training across blocks.
Though knowledge distillation can facilitate information transfer~\cite{hinton2015distilling}, this mutual learning approach can even impede the learning of the earlier blocks when fewer high-memory devices exist. 
Under other ResNet18 experimental settings, ProFL consistently outperforms the baselines, with improvements ranging from 9.2\% to 31.2\% in accuracy.

For the ResNet34 model, its intricate architecture demands more memory from each device to train the full model, further exacerbating memory constraints.
In this scenario, the performance of all algorithms experiences a decline compared to training ResNet18.
Notably, ExclusiveFL becomes completely non-operational in this context, as no device possesses sufficient memory to update the full model.
However, ProFL still demonstrates outstanding performance. Specifically, on the CIFAR10 dataset under the IID condition, ProFL attains a 15.3\% accuracy improvement over AllSmall. 
In contrast to HeteroFL, the absence of devices capable of accommodating the full model training leads to certain channels not being effectively trained, resulting in a significant performance drop of 72.4\%. 
Similarly, DepthFL can only train the front classifiers, and the later classifiers cannot receive effective training. Therefore, when utilizing the ensemble results from all classifiers as the outcome, there is a significant accuracy decrease by 10.5\%. 
Across other ResNet34 experimental settings, ProFL demonstrates accuracy improvements ranging from 18.3\% to 74.2\%.

Regarding the VGG series models, ProFL achieves accuracy improvements of up to 16.1\% on VGG11$\_$bn. Notably, for VGG16$\_$bn, where no device can support the full model training due to its complex architecture, ProFL demonstrates remarkable accuracy gains of up to 82.4\% while concurrently realizing a 57.4\% reduction in peak memory usage.
In conclusion, ProFL consistently delivers superior performance across diverse experimental settings, which can be attributed to three key factors.
First, the progressive training paradigm enables ProFL to build a more complex global model than the largest local model that a device can train.
Second, strategic freezing of converged blocks significantly reduces peak memory usage, thereby allowing more devices with limited memory to participate and achieving a 100\% participation rate. 
Third, we decouple the training process into two stages, designing output modules for each block to assist it in learning the intended feature representation.

\begin{figure*}[!h]
    \centering
    \begin{subfigure}{0.24\textwidth}
        \centering
        \includegraphics[width=\linewidth]{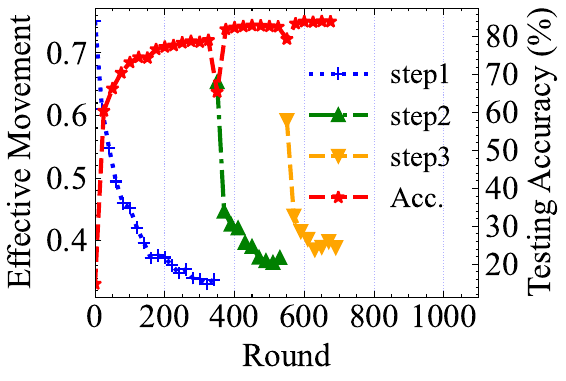}
        \caption{CIFAR10 (IID).}
        \label{perturbation11}
    \end{subfigure}
    \begin{subfigure}{0.24\textwidth}
        \centering
        \includegraphics[width=\linewidth]{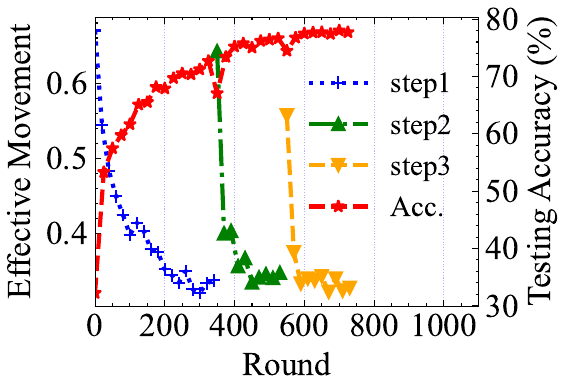}
        \caption{CIFAR10 (Non-IID).}
        \label{perturbation12}
    \end{subfigure}
    \begin{subfigure}{0.24\textwidth}
        \centering
        \includegraphics[width=\linewidth]{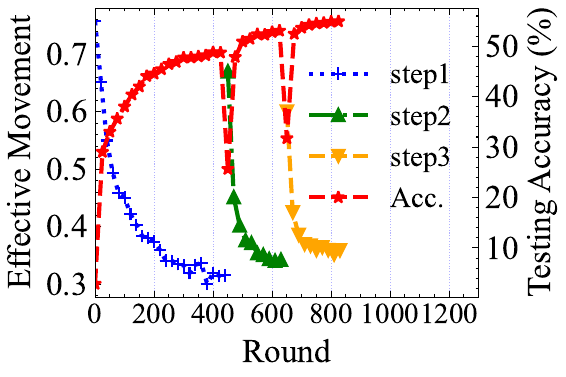}
        \caption{CIFAR100 (IID).}
        \label{perturbation13}
    \end{subfigure}
        \begin{subfigure}{0.24\textwidth}
        \centering
        \includegraphics[width=\linewidth]{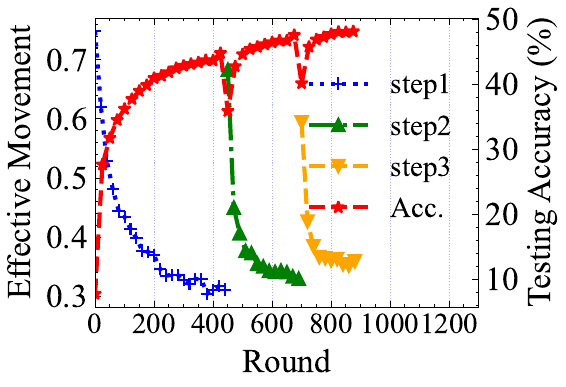}
        \caption{CIFAR100 (Non-IID).}
        \label{perturbation14}
    \end{subfigure}
    \caption{Effective movement serves as a robust indicator reflecting block convergence status in ResNet18. Here, \textit{step} represents the effective movement of the sub-model at each step, while \textit{Acc.} represents the testing accuracy of the corresponding round.}
    \label{perturbation1}
\end{figure*}

\begin{figure*}[!h]
    \centering
    \begin{subfigure}{0.24\textwidth}
        \centering
        \includegraphics[width=\linewidth]{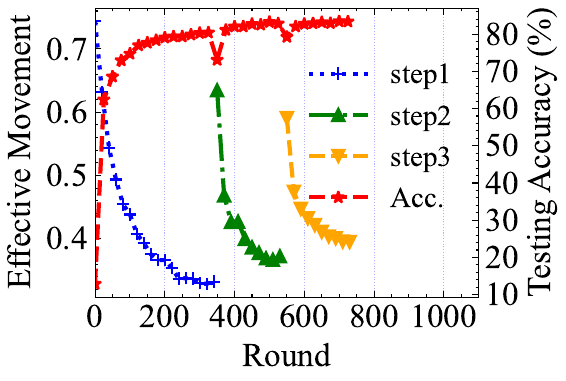}
        \caption{CIFAR10 (IID).}
        \label{perturbation21}
    \end{subfigure}
    \begin{subfigure}{0.24\textwidth}
        \centering
        \includegraphics[width=\linewidth]{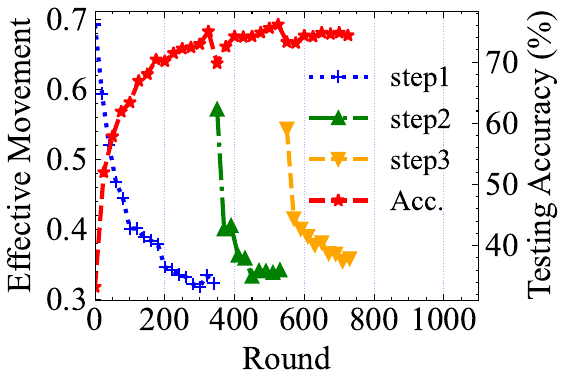}
        \caption{CIFAR10 (Non-IID).}
        \label{perturbation22}
    \end{subfigure}
    \begin{subfigure}{0.24\textwidth}
        \centering
        \includegraphics[width=\linewidth]{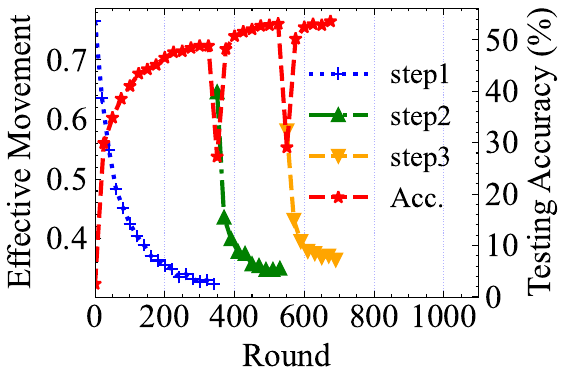}
        \caption{CIFAR100 (IID).}
        \label{perturbation23}
    \end{subfigure}
        \begin{subfigure}{0.24\textwidth}
        \centering
        \includegraphics[width=\linewidth]{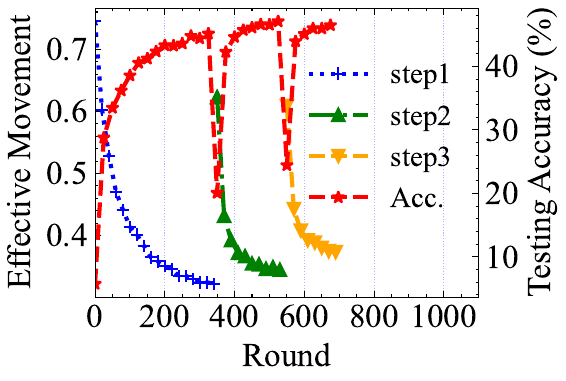}
        \caption{CIFAR100 (Non-IID).}
        \label{perturbation24}
    \end{subfigure}
    \caption{Effective movement serves as a robust indicator reflecting block convergence status in ResNet34. Here, \textit{step} represents the effective movement of the sub-model at each step, while \textit{Acc.} represents the testing accuracy of the corresponding round.}
    \label{perturbation2}

\end{figure*}

\subsection{Ablation Study}\label{sec_ablation}

To validate the effectiveness of the progressive model shrinking and block freezing determination module, we conduct extensive ablation experiments on ResNet18 and ResNet34.

\noindent
\textbf{Effectiveness of Progressive Model Shrinking.} To analyze the impact of progressive model shrinking on the performance of different step-wise sub-models and global models, we enable and disable progressive model shrinking and measure the accuracy of the corresponding models.
We experiment with both IID and Non-IID data distributions on ResNet18 and ResNet34, and the results are shown in Table \ref{tab_shrinking}. Regardless of the data distribution and the global model architecture, it is evident that the progressive model shrinking plays a crucial role in improving the accuracy of both step-wise sub-models and global models. 
Specifically, it improves performance by 0.5\% to 6.7\% for sub-models and 0.9\% to 4.7\% for global models.
The performance improvements stem from two primary factors. First, the parameters trained in the progressive model shrinking stage can be leveraged to initialize the corresponding blocks in the progressive model growing stage, resulting in better optimization results. Second, the basic layers with block-specific information obtained in the progressive model shrinking stage function as output module components, aiding each block in learning the expected feature representation.

\noindent
\textbf{Effectiveness of Block Freezing Determination.} To analyze the effectiveness of the block freezing determination module, we compare it with a parameter-aware method, which allocates different training rounds to blocks proportionally to their parameter counts.
This baseline method is motivated by the empirical observation that blocks with different parameter scales typically demonstrate varying convergence characteristics.
Table \ref{tab_converge1} presents the experimental results, showing that using the block freezing determination module improves performance by 0.8\% to 6.2\% compared to the parameter-aware method.
This superior performance can be attributed to our module's ability to assess block training progress through a more fine-grained scalar perspective, enabling more precise and informed decisions on block freezing timing, which ultimately enhances the model's overall performance.

\begin{table}[!t]
  \centering
   \footnotesize
     \setlength{\tabcolsep}{2pt} 
  \caption{Accuracy of the global model with different freezing methods. $Ours$ represents using the block freezing determination module to control the training pace of each step, and $ParamAware$ represents the allocation of training rounds based on the number of parameters in each block.}
  \label{tab_converge1}
  \resizebox{\linewidth}{!}{

  \begin{tabular}{cccc c}
    \toprule[1pt]
     Distribution &  Dataset & Method&ResNet18 & ResNet34  \\
    \midrule
     \multirow{4}{*}{IID} &\multirow{2}{*}{CIFAR10} & $Ours$  & \textbf{84.1\%} & \textbf{82.2\%}    \\
                & & $ParamAware$ & 81.9\% (-2.2\%) & 81.4\% (-0.8\%)  \\

     &\multirow{2}{*}{CIFAR100} & $Ours$  & \textbf{55.4\%} & \textbf{52.3\%}    \\
                   &  & $ParamAware$  & 50.0\% (-5.4\%) & 49.5\% (-2.8\%)     \\
                                
    \midrule
     \multirow{4}{*}{Non-IID}  & \multirow{2}{*}{CIFAR10}  & $Ours$ & \textbf{78.4\%} & \textbf{74.2\%}    \\
            & &$ParamAware$ & 72.2\% (-6.2\%)& 71.4\% (-2.8\%)   \\

 &\multirow{2}{*}{CIFAR100} & $Ours$  & \textbf{48.3\%}& \textbf{46.1\%}    \\
       &  & $ParamAware$ & 42.8\% (-5.5\%) & 43.0\% (-3.1\%)     \\
    
    \bottomrule[1pt]

  \end{tabular}}
\end{table}

\subsection{Understanding the Effective Movement}

We then analyze how our proposed metric, effective movement, characterizes the learning dynamics of each block.
Figs.~\ref{perturbation1} and \ref{perturbation2} show the experimental results for ResNet18 and ResNet34. The X-axis represents training rounds, while the left and right Y-axes show effective movement and testing accuracy, respectively.
From Fig.~\ref{perturbation11}, it can be observed that effective movement has a high value at the beginning of each step,
signifying that the parameters are rapidly moving towards the optimal point.
Gradually, as gradients decrease and some parameters converge, it decreases until reaching a converged state. This signals that the current block's parameters are near the optimal point, and further updates would offer marginal benefits, allowing for freezing and initiating the training of the next block. Notably, the convergence status of effective movement aligns consistently with accuracy curves in red, reinforcing effective movement as a robust indicator of block training progress. Outliers due to Non-IID may arise, as shown in Fig.~\ref{perturbation2}b. ProFL effectively mitigates their impact by using linear least-squares regression to fit the effective movement.

\begin{figure}[!t]
    \centering
    \begin{subfigure}{0.23\textwidth}
        \centering
        \includegraphics[width=\linewidth]{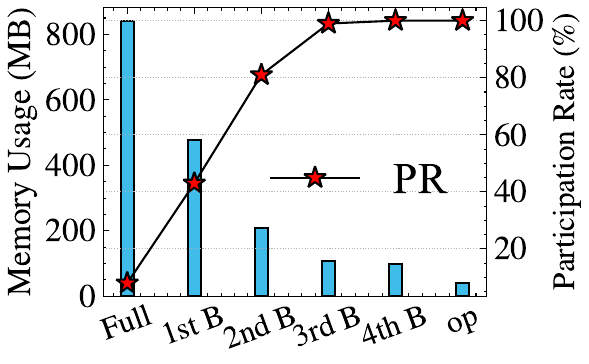}
        \caption{ResNet18.}
        \label{Fig12a}
    \end{subfigure}\hfill
    \begin{subfigure}{0.23\textwidth}
        \centering
        \includegraphics[width=\linewidth]{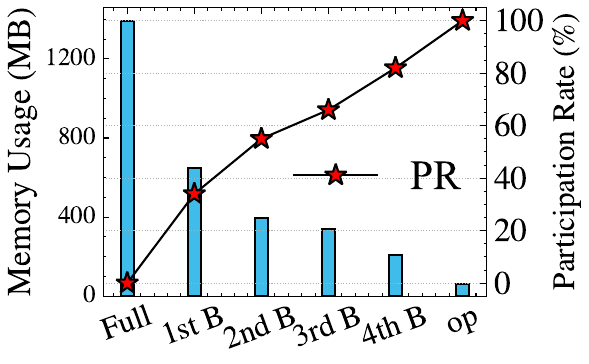}
        \caption{ResNet34.}
        \label{Fig12b}
    \end{subfigure}
    \caption{Memory usage and participation rates when training different blocks. \textit{Full} denotes training the full model, and \textit{op} signifies training the output layer.}
    \label{inclusiveFL}
\end{figure}

\subsection{Understanding the Inclusiveness of ProFL}

To better understand the inclusiveness of ProFL, Fig.~\ref{inclusiveFL} depicts the memory usage and corresponding participation rates of different blocks during the training of ResNet18 and ResNet34. The X-axis illustrates the blocks being trained, the left Y-axis reflects the memory consumption, and the right Y-axis represents the corresponding participation rates. 
We observe that early blocks consume significant memory during local training, primarily due to the large output dimensions of the initial layers, which results in a low participation rate.
As blocks are progressively frozen, memory requirements decrease, allowing more low-memory devices to participate. This progressive nature makes the framework inclusive.

\subsection{Compatibility and Scalability}

We further perform the following experiments to demonstrate the compatibility and scalability of ProFL.
1) We train a ViT~\cite{dosovitskiy2020image} with 12 Transformer encoders on Tiny-ImageNet~\cite{le2015tiny} to demonstrate the compatibility of ProFL with Transformer-based models.
We employ Oracle FL as the baseline, where the global model is trained in an end-to-end manner without considering memory constraints, serving as the theoretical upper bound.
In this case, the Transformer encoder is treated as a basic layer, and the whole model is divided into three blocks in ProFL. Experimental results show that ProFL achieves up to a 3\% accuracy improvement over Oracle FL. This performance gain is attributed to ProFL's progressive block-wise training, which effectively mitigates issues like vanishing gradients that often arise in end-to-end training.
2) We then experiment with the FEMNIST~\cite{caldas2018leaf} dataset, which contains 805,263 images, to demonstrate the scalability of ProFL for large-scale datasets. ResNet18 is used as the global model. The results show that ProFL can achieve 87\% accuracy while reducing peak memory usage by up to 40\%.
3) Moreover, ProFL is compatible with existing FL algorithms, such as FedProx~\cite{li2020federated}, which addresses data heterogeneity. This is because each training round still follows the standard FL process, ensuring seamless integration with these algorithms.

\subsection{Discussion}

\noindent
\textbf{Communication Overhead.} Although dividing the training process into two stages introduces additional communication overhead, this overhead remains moderate and acceptable. This is because, compared to exchanging the full model parameters, we only exchange partial parameters in each training round. Moreover, the parameters obtained in the progressive model shrinking stage are used to initialize each block, accelerating the convergence of the progressive model growing stage. For instance, compared to Oracle FL, ProFL incurs only a 22.1\% increase in communication overhead when training ResNet34 on CIFAR100 under the Non-IID condition. However, it is worth noting that, due to memory limitations, Oracle FL is impractical in real-world scenarios. In contrast, ProFL significantly reduces peak memory usage through its block-wise training paradigm, making it suitable for practical deployment.

\noindent
\textbf{Computation Overhead.} To analyze the computation cost, FLOPs are measured during the training process of ResNet18 on the CIFAR100 dataset. The results demonstrate that the progressive model shrinking stage consumes 0.78 to 1.09$\times$ FLOPs compared with FedAvg~\cite{mcmahan2017communication}, while the progressive model growing stage consumes 0.85 to 1.26$\times$ FLOPs. Moreover, the computation overhead can be further reduced by caching the frozen blocks' activations. This is because, during the training of each block, the preceding blocks remain frozen, with their outputs unchanged. As a result, these blocks and their outputs can be stored on disk. When the device is selected, it only needs to retrieve the activations from disk, thereby effectively avoiding redundant forward computation.
In this way, the overall computation overhead of our ProFL is 1.08 to 1.55$\times$ that of FedAvg.   
Additionally, the computation overhead introduced by the block freezing determination module is negligible. This is because the decision-making process only involves simple calculations between model parameters. For mainstream models on edge devices, the number of model parameters is relatively small, such as MobileNet\_V1 (4.2M). Though introducing extra computing overhead, ProFL effectively breaks the memory wall for FL.

\noindent
\textbf{Latency Analysis.} Although ProFL increases the number of training rounds, the time per round is significantly reduced as only one block is trained within each round. 
Taking the Raspberry Pi 4B training ResNet18 on the CIFAR10 dataset as an example, the training time of the four different blocks is only 26.3\%, 22.9\%, 16.9\%, and 35.9\%, respectively, of the full model training. Thus, the training latency is only 1.57$\times$ that of Oracle FL in total. However, Oracle FL is not feasible for practical applications due to memory limitations. Moreover, the latency can be further reduced by caching the activations of the frozen blocks, allowing the forward propagation process of the frozen blocks to be skipped.

\noindent
\textbf{Block Division.}  Block division can impact training efficiency and memory conservation simultaneously. More blocks can significantly decrease peak memory usage but increase the number of training rounds as fewer layers are grouped into one block. Conversely, fewer blocks reduce memory-saving benefits but enhance training efficiency. 
Thus, when partitioning the model, our primary concerns are preserving the integrity of the original architecture (e.g., skip-connections) and minimizing the number of blocks, all while adhering to memory constraints.
\section{Related Work}

In order to address the resource limitations of the participating devices, several methods have been proposed to customize the local models according to the memory capacity of specific devices. The existing work can be categorized into two approaches: 1) model-heterogeneous FL and 2) partial training.

\noindent
\textbf{Model-Heterogeneous FL.} Model-heterogeneous FL deploys different models on the central server and the corresponding participating devices. The server aggregates these diverse models through knowledge distillation~\cite{hinton2015distilling}. For instance, in FedMD~\cite{li2019fedmd}, devices upload logits on a public dataset to the server once local training is finished. The server then aggregates these uploaded logits and conducts training on the public dataset, facilitating information transfer. Similar approaches include FedDF~\cite{lin2020ensemble}, DS-FL~\cite{itahara2021distillation}, and Fed-ET~\cite{cho2022heterogeneous}.
However, these methods require a high-quality public dataset, which is typically impractical in real-world cases due to privacy issues. Although some algorithms, such as FedZKT~\cite{zhang2022fedzkt}, generate a public dataset using GAN~\cite{goodfellow2020generative}, the training of GAN is often unstable, which deteriorates the global model performance.

\noindent\textbf{Partial Training.} Partial training generates sub-models of varying complexity by scaling the model in terms of width and depth. HeteroFL~\cite{diao2020heterofl} is a representative width scaling method where convolutional layers are scaled to match devices with varying available memory capacity. Similar methods include Federated Dropout~\cite{caldas2018expanding} and FjORD~\cite{horvath2021fjord}. However, these methods may significantly disrupt the integrity of the model's architecture, compromising model performance. Consequently, scaling the model from the depth perspective has been proposed. For instance, DepthFL~\cite{kim2022depthfl} customizes models with different depths based on device memory constraints. However, the global model size in DepthFL is bounded by the device with the largest available memory. In order to break through the limitations, we propose ProFL, a progressive training framework that significantly reduces peak memory footprint during the training process, enabling the participation of low-memory devices.

ProgFed~\cite{wang2022progfed} is the work most closely related to ours, which uses progressive training to reduce communication and computation overhead. 
However, our work differs from it in several key aspects. 1) Different Purposes: ProgFed is designed to accelerate the training process, whereas ProFL is aimed at addressing memory constraints and enhancing model performance. 2) Different Training Paradigms: ProgFed does not freeze blocks, which limits its effectiveness in memory-limited scenarios. In contrast, ProFL reduces memory requirements by intelligently freezing converged blocks. 3) Different Growth Mechanisms: ProgFed employs a fixed growth interval, whereas ProFL integrates a block freezing determination module to decide the optimal timing for model growing. 4) Different Output Modules: ProgFed uses only a linear layer as the output module, which would significantly impair overall performance. To address this issue, we introduce the progressive model shrinking stage to construct corresponding output modules for each block, effectively guiding each block to learn the expected feature representation.

\section{Conclusion}

We present ProFL, a novel progressive training framework to break the memory wall for FL. Specifically, ProFL first divides the global model into blocks based on its original architecture and then trains each block in a progressive manner. In order to preserve the feature representation of each block, we introduce the progressive model shrinking stage, which meticulously constructs corresponding output modules for each block and obtains the initialization model parameters. Moreover, we propose a novel metric from the scalar perspective to control the training pace of each block. Finally, we theoretically prove the convergence of ProFL and conduct extensive experiments to evaluate its effectiveness. The results demonstrate that ProFL effectively reduces peak memory footprint by up to 57.4\% and improves model accuracy by up to 82.4\%.

\begin{acks}
We sincerely appreciate the anonymous shepherd and reviewers for their valuable comments. This research was supported in part by the MYRG-GRG2023-00211-IOTSC-UMDF and SRG2022-00010-IOTSC grants from the University of Macau.
\end{acks}

\bibliographystyle{ACM-Reference-Format}
\balance
\bibliography{reference}

\end{document}